\documentclass[aps,prb,twocolumn,raggedbottom,floatfix]{revtex4}
\usepackage{amsfonts,amssymb,amsmath,mathrsfs,hyperref,bm}
\usepackage{epstopdf}

\usepackage{float}
\usepackage[caption = false]{subfig}
\usepackage[final]{graphicx}
\usepackage{manfnt,pifont}
\newif\ifhyper
\hypertrue    
\ifhyper
\hypersetup{
   citecolor = {red},
   colorlinks = {true}, % false
   linkcolor = {blue},
   urlcolor = {blue} % magenta
}
\fi

\begin{document}

\title{ Phase diagram and thermal Hall conductivity of spin-liquid Kekul\'{e}-Kitaev model  }
	\author{Fatemeh Mirmojarabian}
	\affiliation{Department of Physics, Sharif University of Technology, Tehran 14588-89694, Iran}
	\author{Mehdi Kargarian}
	\email{kargarian@physics.sharif.edu}
	\affiliation{Department of Physics, Sharif University of Technology, Tehran 14588-89694, Iran}
	\author{Abdollah Langari}
	\email{langari@sharif.edu}
	\affiliation{Department of Physics, Sharif University of Technology, Tehran 14588-89694, Iran}
	\date{\today}

\begin{abstract}
 In this work we study the phase diagram of Kekul\'{e}-Kitaev model. The model is defined on a honeycomb lattice with bond dependent  anisotropic exchange interactions making it exactly solvable in terms of Majorana representation of spins in close analogy to the Kitaev model. However, the energy spectrum of Majorana fermions has a multi-band structure characterized by Chern numbers 0, $\pm$1, and $\pm2$. We obtained the phase diagram of the model in the plane of exchange couplings and in the presence of a magnetic field and found chiral topological and trivial spin-liquid ground states. In the absence of magnetic field most part of the phase diagram is a trivial gapped phase continuously connected to an Abelian phase, while in the presence of the magnetic field a topological phase arises. Furthermore, motivated by recent thermal measurements on the spin-liquid candidate $\alpha$-RuCl$_{3}$, we calculated the thermal Hall conductivity at different regimes of parameters and temperatures and found the latter is quantized over a wide range of temperatures. %The robust quantized plateau at $\pi/12$ is the result of nonzero Chern number of the filled bands, which resembles the plateau observed in the measurement of $\alpha$-RuCl$_{3}$ compound of Ref. [\onlinecite{Kasahara}].
 
 %at low temperatures in the chiral phases, we found that the conductivity changes sign as increasing the temperature. We also argue the possible relevance of this model to recent experimental observations.           
 % Motivated by recent measurements of thermal Hall conductivity in the spin-liquid candidate $\alpha$-RuCl$_{3}$,
\end{abstract}


\maketitle
\section{Introduction}
\label{introduction}
In recent years, there has been a surge of interests in strongly correlated Mott insulators with exotic and nontrivial ground states featuring novel states of matter. Of particular interest is the insulating quantum magnets where the strong quantum fluctuations prevent the formation of any long-range magnetic ordering even at zero temperature, the so-called spin-liquids \cite{Lee,Balents}. Despite being a long-sought problem since the original idea proposed by Anderson \cite{ANDERSON}, the experimental realization of spin liquids in materials has remained elusive until the experimental verification of the absence of magnetic ordering in the quasi-two-dimensional organic materials. The organic compounds $\kappa$-(ET)$_2$Cu$_2$(CN)$_3$ \cite{Shimizu,Yamashita} and EtMe$_3$Sb[Pd(dmit)$_2$]$_2$ \cite{Itou,Itou(b),Itou(c)} have triangular-lattice structure and are Mott insulators at ambient pressure with no signature of magnetic ordering, nor anomalies in the specific heat and/or thermal conductivity up to lowest measured mili-Kelvin temperatures. Beside the organic compounds, in the mineral herbertsmithite ZnCu$_3$(OH)$_6$Cl$_2$ with underlying kagome lattice no indication of magnetic ordering was observed at very low temperatures yielding yet another spin-liquid ground state \cite{Helton,Mendels,Han}. The electronic structure of these materials at half-filling is mainly dominated by spin-$1/2$ ions located at the vertices of the underlying lattices. In the Mott phase the underlying low-energy physics can be simply described by the Heisenberg Hamiltonian $H=J_{H}\sum_{<i,j>}\mathbf{S}_{i}\cdot\mathbf{S}_{j}$, where $\mathbf{S}_{i}$ is the spin operator at site $i$ and $J_{H}$ denotes the Heisenberg antiferromagnetic exchange coupling between nearest-neighbor sites. The boson or fermion representation of spins gives rise to a plethora of spin-liquid ground states, gapless or gapped spectrum, and fractionalized excitations, which can partially explain the experimental measurements \cite{Zhou}.

\begin{figure}[t]
\includegraphics[width=0.5\textwidth]{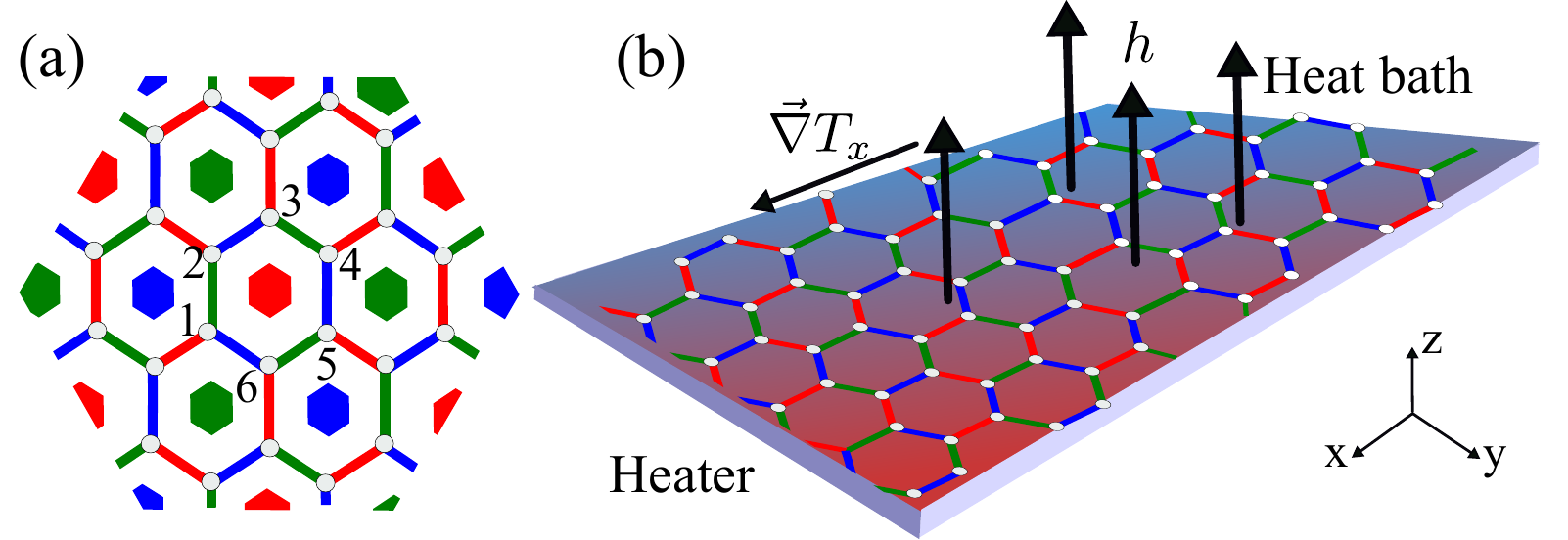}
\caption{(a) The Kekul\'{e}-Kitaev model \cite{Kamfor, Moessner}. Honeycomb 
lattice with six sites in each unit cell. The red, green and blue links 
represent $\sigma^{x}\sigma^{x}$, $\sigma^{y}\sigma^{y}$
and $\sigma^{z}\sigma^{z}$ bonds respectively. (b) A schematic illustration of heat 
conduction to calculate thermal Hall conductivity.}\label{fig:lattice}
\end{figure}

The next generation of two-dimensional magnetic Mott insulators with ground states proximate to a spin-liquid phase arises in materials with 4$d$/5$d$ elements, e.g., the materials containing Ru, Rh, Os, and Ir elements, where the strong spin-orbit coupling manifests large degree of frustration and anisotropic magnetic interactions \cite{Jackeli}. In magnetic iridate compounds (Li, Na)$_2$IrO$_3$ \cite{Singh,Singh(b),Ye,Comin,Takayama,Modic}, the Ir$^{+4}$ ions are located on the vertices of honeycomb lattices stacked along the crystallographic $c$-axis. The low-energy effective Hamiltonian contains the magnetic exchange coupling between the $J_{z}=1/2$ local moments of Ir$^{+4}$ ions, and is described by the Kitaev's model \cite{KITAEV} 
augmented by an isotropic Heisenberg interaction \cite{Doubble}:

\begin{equation}
 H=J_{H}\sum_{<i,j>} \mathbf{S}_{i}\cdot\mathbf{S}_{j}+J_{K}\sum_{<i,j>,\gamma} S^{\gamma}_{i}S^{\gamma}_{j}, 
 \end{equation}
 where the second term with $\gamma=x,y,z$ is anisotropic and bond-dependent, a.k.a, the Kitaev's interactions. Though the model shows a phase transition from a magnetically ordered phase to the Kitaev spin-liquid phase \cite{Doubble} by deceasing $J_{H}$, the inelastic neutron scattering clearly shows an ordered phase at temperatures below $T_N\sim 15$K \cite{Choi}. This observation confirms that in these materials the Heisenberg interaction between magnetic moments is rather strong spoiling the spin-liquid phase. Nevertheless, to understand the underlying zigzag ordered phase, a large degree of anisotropy should be included in the Hamiltonian \cite{Jackeli,Balents,Trebst2017,Lee,Witczak,HwanChun,Reuther,Kargarian}.

The newly discovered ruthenate compound $\alpha$-RuCl$_{3}$ \cite{Banerjee} (and very recently YbCl$_{3}$ \cite{Xing2019}) inspired the realization of the spin-liquid phase, where it turns out the Heisenberg interaction is rather weak and therefore the ground state is possibly proximate to a spin-liquid phase. In the absence of the magnetic field and at low temperatures, i.e., $T\!<\!T_{N}\approx 7$K, the ground state of $\alpha$-RuCl$_{3}$ is characterized by a zigzag antiferromagnetic (AFM) order. The nuclear magnetic resonance
and neutron scattering measurements indicate that the AFM order melts down in a tilted magnetic field applied to the sample when the in-plane component exceeds $\mu _{0 }H^{\ast}_{\parallel} = 7$T, and the spin-liquid phase appears \cite{Kasahara}. The measurements of the 2D thermal Hall conductance show a half-integer quantized plateau at temperatures below $6$K and a possible signature of low-energy fractionalized excitations is demonstrated in microwave absorption measurements \cite{PhysRevB.98.184408}. Thermal transport through the chiral Majorana edge states and the role of bulk phonons discussed in Refs.[\onlinecite{PhysRevX.8.031032,PhysRevLett.121.147201-Balents}] could account for the quantization observed experimentally.   

While a complete understanding of the experimental results still remains to be a far-reaching problem, in most of the theoretical works done so far the focus has mainly been on the original Kitaev model with only two sites in a unite cell leading to a two-band model of Majorana fermions \cite{KITAEV}. In this work we instead consider an alternate of the Kitaev model with a multi-band spectrum, the so-called Kekul\'{e}-Kitaev model \cite{Kamfor, Moessner}. The arrangements of anisotropic bond interactions on the underlying honeycomb lattice is shown in Fig.~\ref{fig:lattice}(a). We first obtain the phase diagram on the latter model. The size of the non-Abelian phase characterized by a finite Chern number does depend on the strength of the time-reversal symmetry-broken perturbation, while in the absence of the latter perturbation most of the phase diagram is characterized by an Abelian model defined on a dual Kagome lattice. Furthermore, we investigate how the multi-band spectrum affects the thermal Hall transport properties. In particular, we show that the thermal Hall conductivity assumes a large quantized value at low temperatures due to the nontrivial band topology of Majorana fermions in the non-Abelian phase. Also, in contrast to the two-band Kitaev model, where the thermal Hall conductivity contribution of the lower band is always positive \cite{motome} (or negative depending on the sign of the applied magnetic field), we found that in the multi-band Kekul\'{e}-Kitaev model the bands contribute with different sings in the thermal Hall conductivity resulting from the Berry curvature profile through the momentum space. The sign change of the thermal Hall conductivity of $\alpha$-RuCl$_{3}$ in a perpendicular magnetic field has been observed experimentally \cite{KasaharaPRL}, an observation which may point toward the necessity of constructing a more realistic multi-band model to understand the physical properties of these materials.                       

The paper is organized as follows. We introduce the Kekul\'{e}-Kitaev model, lattice structure, and its general properties in Sec.\ref{Model}. The effects of time-reversal symmetry breaking and the  phase diagram are discussed in Sec.\ref{threespin}. We then present the results of thermal Hall conductivity in Sec.\ref{Hall}, and Sec. \ref{conclusion} concludes.

\section{ Kekul\'{e}-Kitaev Model and free Majorana Fermion representation\label{Model}}
The exactly solvable spin-$1/2$ Kekul\'{e}-Kitaev model \cite{Moessner, Kamfor} is comprised of two-body
interactions between spins located at the vertices of a honeycomb lattice as   
shown in Fig.~\ref{fig:lattice}(a). 
The spin Hamiltonian of the model is given by
\begin{equation}\label{eq:Hhoneycomb}
H_0=-\sum_{\textless i,j\textgreater,\alpha}J_{\alpha} \sigma_{i}^{\alpha} \sigma_{j}^{\alpha},
\end{equation}
where $\sigma^{\alpha}$ ($\alpha=x, y, z$) denote the Pauli matrices and 
$J_{\alpha}$ are exchange couplings. 
We take $J_{\alpha}>0$ throughout. Note that the model is distinct from the 
famous Kitaev 
model~\cite{KITAEV}, though both
are defined on honeycomb lattice and are exactly solvable via Majorana fermionization as explained
below. In contrast to the Kitaev model, the exchange interactions on the links around the plaquettes are not the same for all cells in the Kekul\'{e}-Kitaev model. 
We use three colors to keep track of the interactions emanating from each vertex. The red, green and blue links
represent $\sigma^{x}\sigma^{x}$, $\sigma^{y}\sigma^{y}$
and $\sigma^{z}\sigma^{z}$ spin interactions, respectively. Now, it is easy to see that we can use
the same colors to label the plaquettes. The color of a plaquette is determined by the color of the outgoing links.
For instance, the red plaquette is the one with red outgoing links and the same holds for blue and 
green plaquettes; see Fig.~\ref{fig:lattice}(a).

\begin{figure*}[t]
\includegraphics[width=1.0\textwidth]{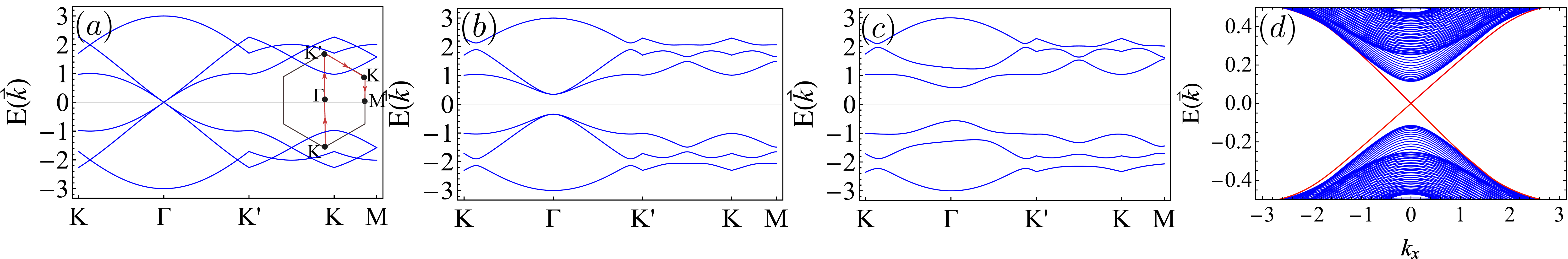}
\caption{(a) Dispersion along high symmetry points for the equal coupling 
strength ($J_{x}=J_{y}=J_{z}$) 
which is four fold degenerate at the $\Gamma$ point. (b) The dispersion 
away from the equal coupling point is gapped. Here 
we considered $J_{x}=1.0$, $J_{y}=0.8$ and $J_{z}=1.2$. (c) The bulk spectrum is also gapped by applying a magnetic field  $h/J_{0}=0.2$ and (d) the  edge states arise due to nontrivial band topology.} 
\label{fig:HSPdispersion}
\end{figure*}

Corresponding to each colored plaquette, we define a plaquette operator which is 
product of Pauli spins located on vertices as follows:

\begin{equation}
W^{{\rm B}}=-\prod_{i=1}^{6}\sigma^{z}_{i},~~
W^{{\rm G}}=-\prod_{i=1}^{6}\sigma^{y}_{i},~~
W^{{\rm R}}=-\prod_{i=1}^{6}\sigma^{x}_{i}.
\end{equation}

These plaquette operators define a set of integral of motions, since they commute with each other $[W^{\gamma}, W^{\gamma'}]=0$ and with the Hamiltonian $[H, W^{\gamma}]=0$, where  $\gamma={\rm R}, {\rm G}, {\rm B}$ (for red, green and blue plaquette). Also, each plaquette operator square identity $(W^{\gamma})^{2}=1$.
Therefore, the Hilbert space of the model is consist of sectors which are eigenspace of 
plaquette operators with eigenvalues $w^{\gamma}=\pm1$.
Analogues to the Kitaev model, in each sector the dimension is still exponentially large calling for a Majorana representation 
of spin operators.

The Majorana fermions obey Clifford algebra, 
$\{c_{i},c_{j}\}=2\delta_{ij}$ and $c_{i}^{2}=1$. 
Following Kitaev \cite{KITAEV} we represent a spin operator by  Majorana fermions $(b^{x},b^{y},b^{z},c)$ as
$\sigma^{\alpha}=\mathrm{i}b^{\alpha}c$ with $\mathrm{i}=\sqrt{-1}$. Hence, the Hamiltonian (\ref{eq:Hhoneycomb}) becomes quadratic
in terms of Majorana operators as

\begin{equation}\label{eq:Hmajorana}
H_0=\frac{\mathrm{i}}{4}\sum_{<i, j>} 2J_{\alpha}u_{i,j}^{\alpha} c_{i}c_{j},
\end{equation} 
where $u_{i,j}^{\alpha}=\mathrm{i}b_{i}^{\alpha}b_{j}^{\alpha}$ is the link operator associated with link ($i$,$j$). 
The latter operators commute with each other $[u_{i,j}^{\alpha},u_{i,j}^{\alpha'}]=0$ and with the Hamiltonian 
$[u_{i,j}^{\alpha},H]=0$, and they square to identity $(u_{i,j}^{\alpha})^{2}=1$ with eigenvalues 
$u_{i,j}^{\alpha}=\pm1$. Thus there is $\mathbb{Z}_{2}$ gauge degrees of freedom on each link. According to the Lieb theorem
\cite{Lieb} the ground state of the model (\ref{eq:Hmajorana}) is in zero-flux sector
corresponding to configuration with $w^{\gamma}=1$ for all plaquettes. Note that $w^{\gamma}$ is defined as product
of link operators around each plaquette $w^{\gamma}=\prod_{(i,j)\in\gamma}u_{i,j}$.
Since $u_{i,j}^{\alpha}=-u_{j,i}^{\alpha}$, to avoid obscurity we select a particular direction for each link. 
We assume that $u_{i,j}^{\alpha}=1$ when the site index $i$ is even and $j$ is odd; see Fig.~\ref{fig:lattice}(a) for site numbering. In the following we work in the zero flux sector with $u_{i,j}^{\alpha}=1$.  

By Fourier transformation to momentum space the Hamiltonian becomes 
\begin{equation}\label{eq:HpureMajorana}
H(\textbf{k})=\frac{\mathrm{i}}{2}\sum_{k}\varPsi_{\textbf{k}}^{T} A(\textbf{k}) \varPsi_{-\textbf{k}},
\end{equation}
where $A(\textbf{k})$ is an antisymmetric matrix given in Appendix \ref{skew}.
%\begin{widetext}
%\begin{equation}\label{eq:AHpureMajorana}
%A(\textbf{k})=
%\left(\begin{array}{cccccc}      
% 0 & -2J_{z}& 0 &-2J_{x} e^{\mathrm{i}k_{x}} & 0 & -2J_{y}\\
%2J_{z}&0&2J_{y}&0&2J_{x} e^{\mathrm{i}(k_{x}-\sqrt{3}k_{y})/2}&0\\
%0 &-2J_{y}&0&-2J_{z}& 0&-2J_{x}e^{-\mathrm{i}(k_{x}+\sqrt{3}k_{y})/2}\\
%2J_{x} e^{-\mathrm{i}k_{x}} &0&2J_{z}&0&2J_{y}&0\\
%0&-2J_{x} e^{-\mathrm{i}(k_{x}-\sqrt{3}k_{y})/2}&0&-2J_{y}&0&-2J_{z}\\
%2J_{y}&0&2J_{x}e^{\mathrm{i}(k_{x}+\sqrt{3}k_{y})/2}&0&2J_{z}&0
%\end{array}\right),
%\end{equation}
%\end{widetext}
and $\varPsi_{\textbf{k}}^{T}=(c_{1\textbf{k}},c_{2\textbf{k}},c_{3\textbf{k}},c_{4\textbf{k}}
,c_{5\textbf{k}},c_{6\textbf{k}}).$

To study the phase diagram we choose a plane in 
parameter space 
($J_{x}+J_{y}+J_{z}=3J_{0}$). We set $J_{0}=1$ as an energy scale. At equal 
coupling strength ($J_{x}=J_{y}=J_{z}=1$) the spectrum is gapless and the 
dispersion is composed of two superimposed Dirac cones at the center of the 
Brillouin zone (BZ); see the bulk spectrum along the high-symmetry lines of BZ 
in Fig.~\ref{fig:HSPdispersion}(a). This is in contrast to the Kitaev model 
\cite{KITAEV}, where the Dirac cones appear at $K$ and $K^{'}$ points. In 
the Kekul\'{e}-Kitaev model the crossing of the Majorana bands occurs at the 
$\Gamma$ point. This has an important consequence on the stability of the 
nodes. While in the former case the model remains gapless until the nodes meet 
at the center of BZ giving rise to a finite region in the phase diagram known 
as B-phase, the latter model is only gapless when all couplings are equal.

\begin{figure*}[t]
\includegraphics[width=0.9\textwidth]{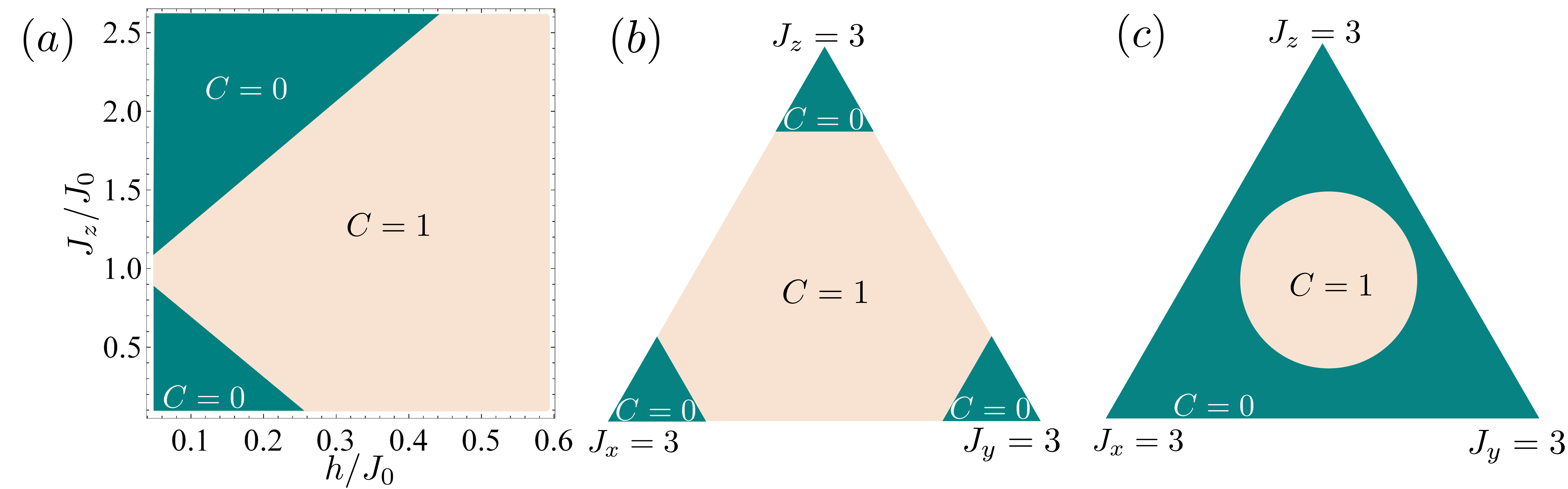}
\caption{The phase diagram of the Kekule-Kitaev model (\ref{eq:Hthreespinintraction}) (a) in $J_z$-$h$ plane, where $J_x\neq J_y\neq J_z$. The light (dark) region corresponds to topological (trivial) phase characterized by Chern number $C=1$ ($C=0$). All phase diagrams are restricted to the plane $J_x+J_y+J_z=3$. The phase diagram is presented at fixed magnetic field for (b) $h=0.4$ and (c) $h=0.2$.}
\label{fig:phasediagram}
\end{figure*}
           
In general there are two ways to open a gap in the spectrum and create a gapped 
spin-liquid phase: (i) making the exchange coupling on one set of bonds, say 
red, to be stronger than the others, or (ii) breaking the time reversal symmetry. 
For the case (i), as shown in Fig.~\ref{fig:HSPdispersion}(b), by a small 
deviation, from equal coupling strength the spectrum becomes gapped. 
The fragile nature of the gapless phase is ascribed to the fact that both nodes 
appear at the same point in BZ, making it susceptible to perturbations, 
which can create the matrix elements between the nodes. In the Kitaev model 
however a finite strength of type (i) is required to move the nodes to the same 
point and then annihilate them. The gapped phase around the gapless point is 
connected to the gapped phase near the 
corner of the phase diagrams without a phase transition, and consequently, they 
should have the same low-energy properties. Near the corners of the phase 
diagram one of the exchange coupling becomes much larger than the others, say 
$J_{z}\gg J_{x}, J_{y}$. This limit is well suited for using the degenerate 
perturbation theory to obtain a low-energy description in terms of the original 
spin degrees of freedom. The effective model becomes a $\mathbb{Z}_{2}$ lattice 
gauge theory defined on the Kagome lattice \cite{Moessner}. The latter lattice 
is obtained by shrinking the blue 
links, corresponding to $J_{z}\sigma^z\sigma^z$ coupling, of the honeycomb 
lattice to effective sites carrying a doublet of pseudospin-1/2 states. 
Therefore the gapped phase in Fig.~\ref{fig:HSPdispersion}(b) is continuously 
connected 
to a phase with Abelian anyon excitations. % Appendix \ref{perturbation}   

\section{breaking the time-reversal symmetry: Chiral spin liquid \label{threespin}}
Now we focus on the case (ii) mentioned in the preceding section to open a gap in the spectrum. This can be 
achieved by applying an external magnetic field $H_{B}=\sum_{i} {\bm B}\cdot{\bm \sigma}_{i}$ to the system, i.e., $H=H_{0}+H_{B}$. We assume that the magnetic field is small. Following Kitaev \cite{KITAEV}, the effect of the magnetic field can be studied perturbatively giving rise to three-spin interaction terms in the Hamiltonian (\ref{eq:Hhoneycomb}) as follows:       

\begin{equation}\label{eq:Hthreespinintraction}
H=-\sum_{\textless i,j\textgreater,\alpha}J_{\alpha} \sigma_{i}^{\alpha} \sigma_{j}^{\alpha}
-h \sum_{i,j,l}\sigma_{i}^{x} \sigma_{j}^{y} \sigma_{l}^{z},
\end{equation} where $h\simeq B^3/\Delta^2$ and we treat it as an independent parameter in the following. Here $\Delta$ is the gap to the excitations of the background fluxes \cite{KITAEV}. Despite having multi-spin interaction terms, the model remains to be exactly solvable. Using the Majorana representation, the above Hamiltonian is rewritten as
\begin{equation}\label{eq:Hmajoranathreebodyterm}
H=\frac{\mathrm{i}}{4}\sum_{<i, j>} 2J_{i,j}^{\alpha}u_{i,j}^{\alpha} 
c_{i}c_{j}+ \mathrm{i} h \sum_{\ll i,j\gg}c_{i}c_{j},
\end{equation}

It is seen that the three-spin term translates to second-neighbor hopping for 
Majorana fermions, and the Hamiltonian retains its bilinear form in fermion 
operators. In momentum space a Bloch Hamiltonian similar to (\ref{eq:HpureMajorana}) is obtained where the antisymmetric matrix is replaced with $A(\mathbf{k})+B(\mathbf{k})$, and the expression for 
$B(\mathbf{k})$ is given in Appendix \ref{skew}.    

The band structures for $h=0.2$ is 
shown in Fig.~\ref{fig:HSPdispersion}(c). The spectrum becomes fully gapped 
throughout the BZ. We shall discuss that this gapped phase is distinct from the gapped 
phase in  Fig.~\ref{fig:HSPdispersion}(b). The distinction can be made more explicit and quantitative by evaluating the 
 first Chern number
\begin{equation} \label{chern}
 C_{n}=\frac{1}{2\pi}\int_{\mathrm{BZ}} d\bold{k}~ \Omega^{z}_{n}(\bold{k}),
\end{equation}
where $\Omega^z_{n}(\bold{k})$ is the Berry curvature:
$\boldsymbol{\Omega}_{n}(\bold{k})=\mathrm{i} \langle 
\boldsymbol{\nabla}_{\mathbf{k}} u_{n}| \times 
|\boldsymbol{\nabla}_{\mathbf{k}} u_{n}\rangle$ with $\arrowvert 
u_{n}(\bold{k})\rangle$ as the periodic part of the Bloch wave function in the 
$n$-th band with energy dispersion $\varepsilon_{n\bold{k}}$, i.e., 
$H(\mathbf{k})\arrowvert 
u_{n}(\bold{k})\rangle=\varepsilon_{n\bold{k}}\arrowvert u_{n}(\bold{k})\rangle$. The integration is taken over the entire BZ. %We use the gauge-invariant method developed in Ref.\cite{Fukui} to compute the Chern number (\ref{chern}) for all bands.     

Lets take $J_x=J_y=J_z=J_0$ and $h/J_0=0.2$ for the moment. The evaluation of the Chern number 
shows that the band structure shown in Fig.~\ref{fig:HSPdispersion}(c) is 
topologically nontrivial: the Chern numbers read as $(0,-1,2,-2,1,0)$ for the 
bands from lowest to highest energies. Hence the occupied Bloch bundle, the three occupied bands 
corresponding to half-filling, carries a total Chern number of $+1$. This 
finding immediately implies that the model should carry gapless edge states 
along the one-dimensional boundary. We diagonalize the Hamiltonian 
(\ref{eq:Hmajoranathreebodyterm}) in a ribbon geometry, where the spectrum is 
shown in Fig.~\ref{fig:HSPdispersion}(d). It's clearly seen that the chiral 
edge modes cross the bulk band gap due to the topological bulk Bloch bands. The band 
structure is however trivial in regions far away from $J_x=J_y=J_z=J_0$ point in the parameter space  
and with small $h$ as characterized by the Chern numbers as $(0,1,-1,1,-1,0)$ 
yielding occupied bands 
with total zero Chern number. 

Having obtained a simple picture of the band structure for a few representative 
points in the parameter space, we now present the full phase diagram of the free Majorana model 
(\ref{eq:Hmajoranathreebodyterm}). We obtained two types of phase diagram. Fist we tune the $J_{z}$ and $h$ parameters across a wide range of values, and the obtained phase diagram is shown in Fig.~\ref{fig:phasediagram}(a). The region 
with total Chern number $C=1$, as explained above, has the Chern number 
$(0,-1,2,-2,1,0)$ for the Bloch bands. As we shall discuss in the next section it would have important 
consequences for the thermal Hall conductivity at low fields. The region with $C=0$ 
is trivial with Chern number distribution for all band as $(0,1,-1,1,-1,0)$. Second, we obtained a phase diagram in $J_x+J_y+J_z=3$ plane at two values of magnetic field $h=0.4$ and $h=0.2$ as shown, respectively, in Fig.~\ref{fig:phasediagram}(b) and Fig.~\ref{fig:phasediagram}(c). For larger value of $h$ the majority part of the phase diagram is occupied by the topologically nontrivial phase with $C=1$. By decreasing the magnetic field this region shrinks to a smaller one around the isotropic point.

\section{Thermal Hall conductivity \label{Hall}}
In the preceding section we obtained the phase diagram of the multi-band Majorana model 
(\ref{eq:Hmajoranathreebodyterm}) consisting of topological and trivial phases. In this section we want to see what are the implications of these 
phases and the phase transition between them on the outcomes of the 
experimental probes. A natural consequence of the former phase is the existence of gapless chiral states propagating along the edges of the system. Since the edge mode is chiral and topologically protected, a sort of quantization is expected to occur in appropriate measurements. Since the low-energy properties of the model are described 
by Majorana fermions, which are neutral particles, there is no charge response 
in the system. Yet, the thermal probes can measure the response of Majorana 
fermions as they can carry energy and consequently heat through a system 
subjected to a thermal gradient $\boldsymbol{\nabla}_xT$, where $T$ is the 
temperature. A sketch of the measurement is shown in Fig.~\ref{fig:lattice}(b) 
in a close analogy with the set up used in recents experiments on 
$\alpha$-RuCl$_{3}$ \cite{Kasahara, KasaharaPRL}. 

Of particular interest for our study of topological phases is to evaluate the thermal Hall conductivity \cite{Go2018}, which measures the 
transverse heat current $J^{Q}_{y}=-\kappa_{xy}(\boldsymbol{\nabla}_xT)$. The expression 
for $\kappa_{xy}$ is as follows:  
\begin{equation} \label{eq:thermalHall}
\kappa_{xy}=\frac{-k_{B}^{2}}{\hbar 
\mathcal{A}T} \int d\epsilon \epsilon^{2} \frac{\partial f(\epsilon, 
T)}{\partial \epsilon}
\sum_{\bold{k},n}\Omega^{z}_{n}(\bold{k})
\end{equation}
where $\mathcal{A}$ is the area of the system, $k_{B}$ and $\hbar$ are the 
Boltzmann and the reduced Planck constants, respectively. We set 
$k_{B}=\hbar=1$ in the following and restore when needed. Here $f$ is the Fermi-Dirac distribution 
function of the $n$-th band. The summation runs over the first BZ. 

\begin{figure}[t]
\includegraphics[width=0.45\textwidth]{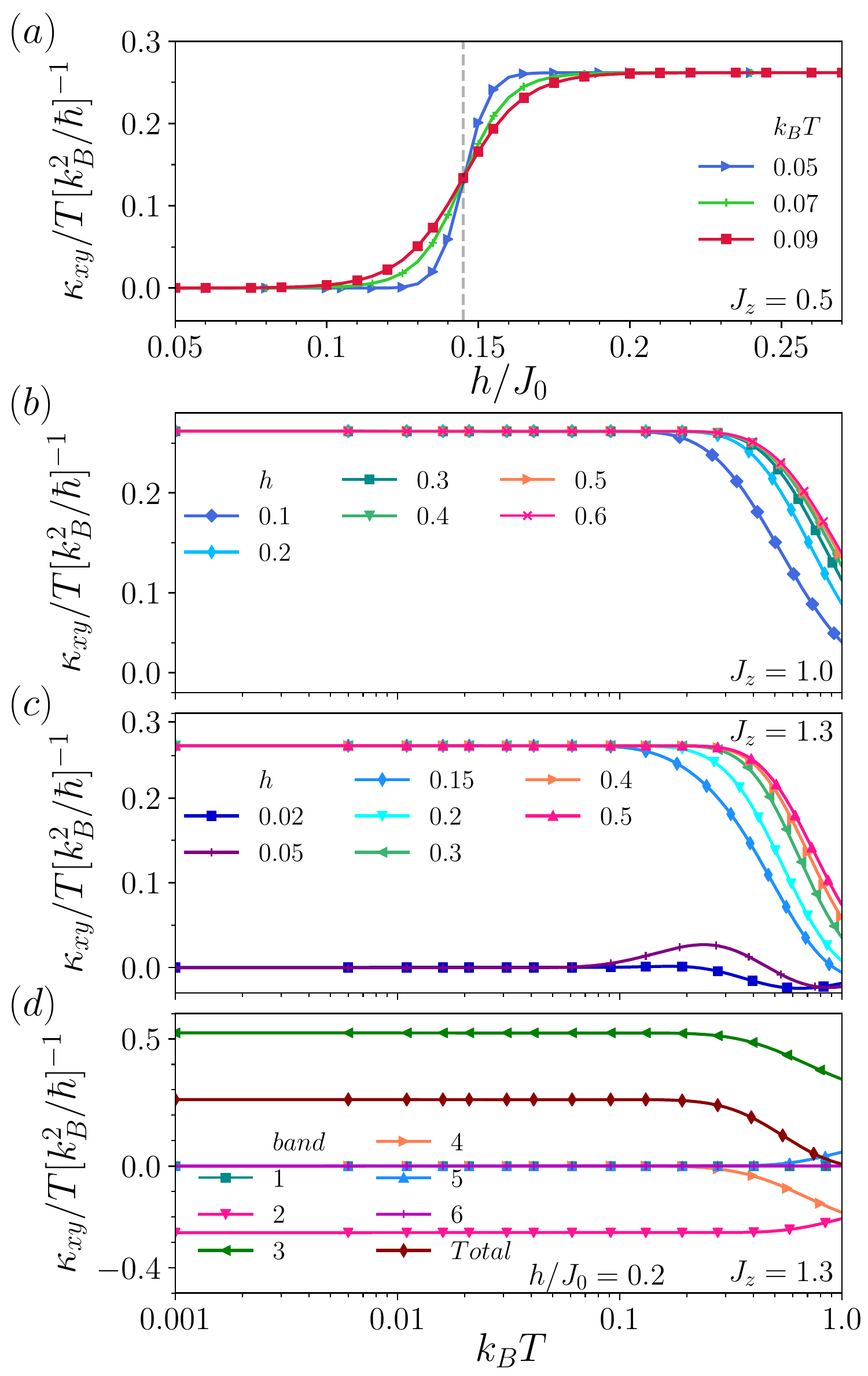}
\caption{ The variation of $\kappa_{xy}/T$ (a) across a topological phase transition at different low temperatures, (b) versus temperature at various magnetic fields at $J_x=J_y=J_z=1$ and (c) away from the equal coupling exchanges, (d) the contributions of individual bands to thermal Hall conductivity. In all plots the saturation of $\kappa_{xy}/T$ at quantized values has a topological origin as discussed in main text.}
\label{fig:kxy}
\end{figure}

The results of $\kappa_{xy}/T$ for various cases are shown in 
Fig.~\ref{fig:kxy}. We begin by calculating the thermal Hall conductivity along a particular cut in the phase diagram Fig.~\ref{fig:phasediagram}(a). We set $J_{z}=0.5$ and plot $\kappa_{xy}/T$ versus the magnetic field in Fig.~\ref{fig:kxy}(a) at different temperatures. Note that in these plots we restored $\hbar$ and $k_B$. At low temperatures the value of $\kappa_{xy}/T$ in the trivial phase with $C=0$ is nearly zero and a great enhancement is observed across the topological phase transition around $h/J_0\simeq 0.15$. The striking feature is that the value of $\kappa_{xy}/T$ saturates to a  plateau quantized at $\pi/12$ as also expected from the number of chiral boundary mode. At hight temperatures the increment around the phase transition is slightly smeared out, yet the quantization remains intact away from the transition.          

Next we study the variation of the thermal Hall conductivity with temperature in both phases. First we consider the case with $J_x=J_y=J_z=1$, where the model is gapless in the absence of the magnetic field. As discussed in the preceding section a finite field opens a gap and the system immediately runs into a topological phase. In this phase the behavior of $\kappa_{xy}/T$ with temperature at different fields is shown in Fig.~\ref{fig:kxy}(b). A clear observation is that a robust quantized value of $\kappa_{xy}/T$ at $\pi/12$ occurs at a wide range of temperatures $T< 0.2$. At higher temperatures there is strong deviation from the quantized value. Indeed at the high temperatures the high energy band are thermally occupied by the Majorana fermions and consequently the contributions from all bands gives rise to a temperature dependent value. Note that at very high temperatures the $\kappa_{xy}$ in (\ref{eq:thermalHall}) is proportional to $\sum_{\bold{k},n}\Omega^{z}_{n}(\bold{k})$ over all bands which vanishes. 
 
Fig.~\ref{fig:kxy}(c) shows the same plot of $\kappa_{xy}/T$ in the Abelian phase with $J_z=1.3$. At small magnetic field where $C=0$ the $\kappa_{xy}/T$ vanishes at low temperatures. A hump in $\kappa_{xy}/T$ is observed at temperatures around $T\simeq 0.2$, which is likely due to the thermal occupation of bands with finite Chern number right above the gap. When the strength of the field is increased, a pronounced increment is observed in $\kappa_{xy}/T$ at low temperatures, which is again quantized to the value of $\pi/12$ akin to the nontrivial band topology with $C=1$. Finally, we diagnose the contribution of different bands to quantized plateau of $\kappa_{xy}/T$. To do so, in Fig.~\ref{fig:kxy}(d) we plot $\kappa_{xy}/T$ for all six bands along with the total one at fixed $h=0.2$. For this field the Chern numbers for all bands are as $(0,-1,2,-2,1,0)$ from the lowest to the highest ones. While the lowest occupied band gives no contribution, the second occupied band gives a plateau at $-\pi/12$ resulting from a band with Chern number $-1$. The third occupied band yields a plateau at $\pi/6$ due to the band with Chern number $+2$. Indeed the total contribution is quantized to $\pi/12$ from all bands.

\section{Summary and Conclusions \label{conclusion}}
In this work we have studied the Kekul\'{e}-Kitaev model \cite{Kamfor, Moessner} whose spectrum is given by a multi-band model of Majorana fermions in terms of exchange couplings $J_x$, $J_y$, $J_z$, and a magnetic field $h$ as time-reversal breaking perturbation. Our main findings can be summarized as follows: we (i) found that at $J_x=J_y=J_z$ and $h=0$ the spectrum is gapless and a gapped phase arises away from $J_x=J_y=J_z$ point continuously connected to an Abelian phase whose low-energy spectrum is given by abelian anyons on the Kagome lattice, (ii) obtained the full phase diagram of the model in the presence of a magnetic field and established that the magnetic field drives the system through the topological and trivial phases characterized by total integer Chern numbers $\pm1$ and $0$ of occupied bands, respectively, (iii) systematically evaluated the field and temperature dependences of the thermal Hall conductivity and found that it shows distinct behaviors in topological and trivial phases, and (iv) observed a quantized plateau at low temperatures. The latter quantization is a resemblance of half-quantized plateau observed recently in thermal Hall measurements in compound $\alpha$-RuCl$_{3}$ \cite{Kasahara}. Our results may suggest that the multi-band Kekul\'{e}-Kitaev model can also be considered as an alternative model and perhaps, when supplemented with other isotropic and anisotropic interactions, to describe other aspects of the experimental observations such as the sign change of thermal Hall conductivity, which we leave it for future study.

\section{Acknowledgements}
The authors would like to acknowledge the support from Sharif University
of Technology under Grant No. G690208.
\appendix
\section{Antisymmetric skew matrices \label{skew}}
In this appendix we present the full expression of skew antisymmetric matrices $A(\mathbf{k})$ and $B(\mathbf{k})$ appearing in the Bloch Hamiltonian on Majorana fermions. Let us assume that the primitive unite vectors of the honeycomb lattice are $\textbf{a}_{1}=(1,0)$ and $\textbf{a}_{2}=(1/2,\sqrt{3}/2)$. The matrices are as follows: 

\begin{widetext}
\begin{equation}\label{eq:AHpureMajorana}
A(\textbf{k})=
\left(\begin{array}{cccccc}      
0 & -J_{z}& 0 &-J_{x} e^{\mathrm{i}\textbf{k}\cdot\textbf{a}_{1}} & 0 & 
-J_{y}\\
J_{z}&0&J_{y}&0&J_{x} 
e^{\mathrm{i}\textbf{k}\cdot(\textbf{a}_{1}-\textbf{a}_{2})}&0\\
0 &-J_{y}&0&-J_{z}& 0&-J_{x}e^{-\mathrm{i}\textbf{k}\cdot\textbf{a}_{2}}\\
J_{x}  e^{-\textbf{k}\cdot\textbf{a}_{1}} &0&J_{z}&0&J_{y}&0\\
0&-J_{x} 
e^{-\mathrm{i}\textbf{k}\cdot(\textbf{a}_{1}-\textbf{a}_{2})}&0&-J_{y}&0&-J_{z}\\
J_{y}&0&J_{x}e^{\mathrm{i}\textbf{k}\cdot\textbf{a}_{2}}&0&J_{z}&0
\end{array}\right),
\end{equation}

\begin{eqnarray} \label{eq:Bk}
\nonumber B(\textbf{k})=&&
h\left(\begin{array}{cccccc}      
0 & 0& -1 &0&1&0\\
0&0&0&-1&0&1\\
1&0&0&0&-1&0\\
0 &1&0&0&0&-1\\
-1&0&1&0&0&0\\
0&-1&0&1&0&0
\end{array}\right)+
\nonumber	h\left(\begin{array}{cccccc}      
0 & 0& -e^{\mathrm{i}\textbf{k}\cdot\textbf{a}_{1}} 
&0&e^{\mathrm{i}\textbf{k}\cdot\textbf{a}_{1}}&0\\
0&0&0&-e^{\mathrm{i}\textbf{k}\cdot\textbf{a}_{1}}&0&0\\
e^{-\mathrm{i}\textbf{k}\cdot\textbf{a}_{1}}&0&0&0&0&0\\
0&e^{-\mathrm{i}\textbf{k}\cdot\textbf{a}_{1}}&0&0&0&-e^{-\mathrm{i}\textbf{k}\cdot\textbf{a}_{1}}\\
-e^{-\mathrm{i}\textbf{k}\cdot\textbf{a}_{1}}&0&0&0&0&0\\
0&0&0&e^{\mathrm{i}\textbf{k}\cdot\textbf{a}_{1}}&0&0
\end{array}\right)\\&&
\nonumber +h\left(\begin{array}{cccccc}      
0 & 0&0 &0&e^{\mathrm{i}\textbf{k}\cdot(\textbf{a}_{1}-\textbf{a}_{2})}&0\\
0&0&0&-e^{\mathrm{i}\textbf{k}\cdot(\textbf{a}_{1}-\textbf{a}_{2})}&0&e^{\mathrm{i}\textbf{k}\cdot(\textbf{a}_{1}-\textbf{a}_{2})}\\
0&0&0&0&-e^{\mathrm{i}\textbf{k}\cdot(\textbf{a}_{1}-\textbf{a}_{2})}&0\\
0 &e^{-\mathrm{i}\textbf{k}\cdot(\textbf{a}_{1}-\textbf{a}_{2})}&0&0&0&0\\
-e^{-\mathrm{i}\textbf{k}\cdot(\textbf{a}_{1}-\textbf{a}_{2})}&0&e^{-\mathrm{i}\textbf{k}\cdot(\textbf{a}_{1}-\textbf{a}_{2})}&0&0&0\\
0&-e^{-\mathrm{i}\textbf{k}\cdot(\textbf{a}_{1}-\textbf{a}_{2})}&0&0&0&0
\end{array}\right)\\&&
+h\left(\begin{array}{cccccc}      
0 & 0& -e^{\mathrm{i}\textbf{k}\cdot\textbf{a}_{2}} &0&0&0\\
0&0&0&0&0&e^{-\mathrm{i}\textbf{k}\cdot\textbf{a}_{2}}\\
e^{-\mathrm{i}\textbf{k}\cdot\textbf{a}_{2}}&0&0&0&-e^{-\mathrm{i}\textbf{k}\cdot\textbf{a}_{2}}&0\\
0 &0&0&0&0&-e^{-\mathrm{i}\textbf{k}\cdot\textbf{a}_{2}}\\
0&0&e^{\mathrm{i}\textbf{k}\cdot\textbf{a}_{2}}&0&0&0\\
0&-e^{\mathrm{i}\textbf{k}\cdot\textbf{a}_{2}}&0&e^{\mathrm{i}\textbf{k}\cdot\textbf{a}_{2}}&0&0
\end{array}\right).	
\end{eqnarray}
\end{widetext}	
	
%\bibliography{refs}


\end{document}